\documentclass[prd,amsmath,amssymb,superscriptaddress,floatfix,nofootinbib,10pt]{revtex4}
\usepackage{times}
\usepackage{amssymb,amsbsy,amsmath,amsfonts}
\usepackage{graphicx}
\usepackage{float}
\usepackage{color}
\usepackage{morefloats}
\usepackage{rotating}
\usepackage{srcltx}
\usepackage{slashed}
\usepackage{multirow}
\usepackage{verbatim}
\usepackage{hyperref}
\usepackage{tabularx}
\usepackage{bm}
\allowdisplaybreaks[4]

\usepackage{bbding}
\usepackage{threeparttable}
\usepackage{soul}

\DeclareUnicodeCharacter{2212}{\ensuremath{-}}

\newcommand{\PreserveBackslash}[1]{\let\temp=\\#1\let\\=\temp}
\newcolumntype{C}[1]{>{\PreserveBackslash\centering}p{#1}}
\newcolumntype{R}[1]{>{\PreserveBackslash\raggedleft}p{#1}}
\newcolumntype{L}[1]{>{\PreserveBackslash\raggedright}p{#1}}

\newcommand{\be}{\begin{equation}}
\newcommand{\ee}{\end{equation}}

\begin{document}

\title{ \boldmath Study of possible $DND^*$ bound states}

\author{Victor Montesinos}
\email[]{vicmonte@ific.uv.es}
\affiliation{Departamento de Física Teórica and IFIC, Centro Mixto Universidad de Valencia-CSIC Institutos de Investigación de Paterna, 46071 Valencia, Spain}

\author{Jing Song}
\email[]{Song-Jing@buaa.edu.cn}
\affiliation{School of Physics, Beihang University, Beijing, 102206, China}
\affiliation{Departamento de Física Teórica and IFIC, Centro Mixto Universidad de Valencia-CSIC Institutos de Investigación de Paterna, 46071 Valencia, Spain}

\author{Wei-Hong Liang}
\email[]{liangwh@gxnu.edu.cn}
\affiliation{Department of Physics, Guangxi Normal University, Guilin 541004, China}
\affiliation{Guangxi Key Laboratory of Nuclear Physics and Technology,
Guangxi Normal University, Guilin 541004, China}

\author{ Eulogio Oset}
\email[]{oset@ific.uv.es}
\affiliation{Departamento de Física Teórica and IFIC, Centro Mixto Universidad de Valencia-CSIC Institutos de Investigación de Paterna, 46071 Valencia, Spain}

\author{ Juan Nieves}
\email[]{ jmnieves@ific.uv.es}
\affiliation{Departamento de Física Teórica and IFIC, Centro Mixto Universidad de Valencia-CSIC Institutos de Investigación de Paterna, 46071 Valencia, Spain}

\author{ Miguel Albaladejo}
\email[]{ Miguel.Albaladejo@ific.uv.es}
\affiliation{Departamento de Física Teórica and IFIC, Centro Mixto Universidad de Valencia-CSIC Institutos de Investigación de Paterna, 46071 Valencia, Spain}

\begin{abstract}
  We start from a recently favored picture in which the $\Lambda_c(2940)$ and $\Lambda_c(2910)$ correspond mostly to $ND^*$ bound states with $J^P = 1/2^-,\, 3/2^-$ and then add a $D$ as a third particle, looking for the possible binding of the $DND^*$ three body system within the framework of the Fixed Center Approximation. We find that the system is bound with respect to the corresponding $\Lambda_c^* D$ threshold with a binding of about $60$ MeV and a width of about $90$ MeV. Alternatively we assume a cluster of $ND$ and a $D^*$ meson interacting with the cluster and we find similar results. The observation of these states of $J^P = 1/2^+,\, 3/2^+$ would provide new and valuable information concerning the $DN$ and $D^* N$ interaction, an issue of current debate.
\end{abstract}

%\pacs{13.75.Ev,12.39.Fe,21.30.Fe}
%\keywords{}

%\date{\today}
%\author{Jing Song and Victor}

\maketitle

\section{introduction}
The three-body bound states are receiving more and more attention and a large list of them can be seen in the review paper~\cite{MartinezTorres:2020hus} (see Table 1 of that work). 
Another relevant review is given in Ref.~\cite{Liu:2024uxn} proposing different methods to use the experimental data to obtain information of the nature of the states.
A complementary view can be seen in the work of Ref.~\cite{Wu:2022ftm}, where the point is stressed that, while there is no meson number conservation, which favors decays of multimeson states in states with fewer mesons, flavor is conserved in strong interactions, which makes multimeson states with different flavors certainly more stable. Another stabilizing factor is high spin, which leads to decay modes with high angular momentum which are not favored. In this direction, multirho states~\cite{Roca:2010tf}  and a $K^*$ plus multirho states of high spin~\cite{Yamagata-Sekihara:2010muv} were investigated and found compatible with existing states. 
In particular, states containing one nucleon and two mesons have been the object of special attention. The  $\pi\pi N$ system has been investigated in Ref.~\cite{Desplanques:2008fg}, the $K \bar{K} N$ in Refs.~\cite{MartinezTorres:2010zv,Xie:2010ig,Jido:2008kp},
and the $N D K, ~N D \bar{K}, ~N D  \bar{D}$ in Ref.~\cite{Xiao:2011rc}. The $N D  \bar{D}$ system is shown to the bound in Ref.~\cite{Xiao:2011rc}, which is not surprising since the $D \bar{D}$ interaction is rather strong, leading to a bound   $D \bar{D}$ state~\cite{Gamermann:2006nm,Nieves:2012tt,Hidalgo-Duque:2012rqv,Prelovsek:2020eiw,Deineka:2021aeu}, and the $DN$ interaction is also attractive as we will see. 
In between, the big surprise of the finding of the $T_{cc}$~\cite{LHCb:2021vvq,LHCb:2021auc},
identifying it with a $DD^*$ molecular state ~\cite{Ling:2021bir,Dong:2021bvy,Feijoo:2021ppq,Fleming:2021wmk,Ren:2021dsi,Chen:2021cfl,Albaladejo:2021vln,Du:2021zzh,Baru:2021ldu,Santowsky:2021bhy,Deng:2021gnb,Ke:2021rxd,Agaev:2022ast,Kamiya:2022thy,Isar:2023ugc,Abreu:2022sra,Chen:2022vpo,Albaladejo:2022sux,Peng:2021hkr,Ortega:2022efc}, poses the challenge to investigate whether the system $DND^*$ could be bound. The challenge was soon taken and the system was studied in Ref.~\cite{Luo:2022cun} using the Gaussian expansion method.\footnote{As usual, there are also proposals as compact tetraquark
objects~\cite{Carlson:1987hh,Silvestre-Brac:1993zem,Semay:1994ht,Pepin:1996id,Ballot:1983iv,Zouzou:1986qh,Guo:2021yws,Xin:2021wcr}. However, a thorough study done in Ref.~\cite{Dai:2023kwv}
concludes that, while from the pole position one cannot rule out the compact tetraquark picture, the predicted scattering lengths and effective range assuming that structure are in sheer disagreement with present data.}

The work of Ref.~\cite{Luo:2022cun}  constructs potentials for the $D N,~ D^{*}N$ and $D D^{*}$ interaction based on the one boson exchange model, and then use the Gaussian expansion method~\cite{Hiyama:2012sma,Hiyama:2003cu} to obtain the binding of the $D N D^{*} $ system. The conclusion is that the system admits bound states solutions in $I(J^P)=1/2\left(1 / 2^{+}\right)$ and $1/2\left(3 / 2^{+}\right)$, but the binding energies depend strongly on the cutoff assumed in the regularization of the loops and other parameters of the model.
As usual in this variational method, the widths of the states are not evaluated.

In the present work we retake the idea from a different perspective using the fixed center approximation (FCA) to the Faddeev equations as the framework. The latter, although less accurate in principle, relies nevertheless on empirical information that constrains the output and at the same time provides the width of the states~\cite{MartinezTorres:2020hus,Foldy:1945zz,Brueckner:1953zz}. In the FCA there is a cluster of two bound particles and a third particle that scatters with them. The FCA also assumes that the bound cluster is very stable and  is not changed by the scattering of the external particle with its components. This is favored by having the cluster to be very bound and the external particle lighter than the components of the cluster. One could think of a cluster of $DD^*$ (the $T_{cc}$) and the nucleon external particle. Yet, we discard this structure  because the $0.36~\mathrm{MeV}$ binding energy is too small to think that the $T_{cc}$ would not be broken by the scattering with the nucleon. The $D N$ and $D^{*} N$ systems are far more bound in the different models and this makes these systems better suited to play the role of the cluster in the FCA. We choose the heavier cluster of this type and then we have $D^* N$ as a cluster and $D$ as the external particle. We shall also investigate the related system with a $D N$ cluster and $D^*$ as the
external particle.

In the first scenario we will need the  $D N,~ D D^*$  interaction and the wave function of the $D^* N$, and  this leads us to the following discussion. The $DN$ interaction has been the subject of intense study. In Refs.~\cite{Tolos:2004yg,Tolos:2005ft} it was considered by analogy to the $\bar{K} N$ interaction. More accurate calculations were done in Refs.~\cite{Hofmann:2005sw,Lutz:2005vx,Mizutani:2006vq,Tolos:2007vh,Molina:2009zeg} considering all coupled channels to $D N$, allowing them to interact using vector exchange, the explicit picture that leads to chiral Lagrangians in SU(3)~\cite{Ecker:1989yg} (see practical examples in Appendix A of Ref.~\cite{Dias:2021upl}). In these works, the free parameters of the theory, basically the cutoffs regulating the loops,  are fitted assuming that the interaction generates the $\Lambda_c(2595)$ resonance. This view was challenged in Ref.~\cite{Garcia-Recio:2008rjt} using  a model with meson exchange where the couplings satisfy SU(8) spin-flavor symmetry,\footnote{This approach is based on a consistent $SU(6)_\mathrm{lsf} \times HQSS$ extension of the Weinberg–Tomozawa (WT) $\pi N$ interaction, where “lsf” stands for light quark–spin–flavor symmetry, and HQSS for heavy quark spin symmetry.} that mixes pseudoscalar-baryon and vector-baryon channels. There it was found that the  $\Lambda_c(2595)$ has also a strong coupling to  $D^{*} N$, and that this latter channel gives rise predominantly to the  $\Lambda_c(2625)$. Similar results are reported in Ref.~\cite{Romanets:2012hm}. The issue of the $\Lambda_c(2625)$ ($3/2^-$) being the spin partner of the $\Lambda_c(2595)$ ($1/2^-$) is retaken in  Ref.~\cite{Du:2022rbf} using elements of heavy quark symmetry and QCD lattice results, showing that there could be strong deviations from that picture. The study of the $T_{cc}$ and $T_{\bar{c}\bar{c}}$ mesons in matter is studied along these lines in Ref.~\cite{Montesinos:2023qbx} as a means of getting further information on the subject.

There are strong arguments to  think that the $\Lambda_c(2595)$ is not predominantly a molecular state of $DN$, $\pi \Sigma_c$ and  other channels, in analogy to the  $\Lambda(1405)$. Indeed, while the higher $\Lambda(1405)$ state, $\Lambda(1420)$, is only  10~MeV below the  $\bar{K}N$ channel, the $\Lambda_c(2595)$  is about 200~MeV below the $DN$ channel. In the region of the $\Lambda_c(2595)$  there are states of the constituent quark model close by~\cite{Yoshida:2015tia} which can account for it, or at least in a fair amount. This discussion is made in Ref.~\cite{Nieves:2019nol}  and elaborated further in Ref.~\cite{Nieves:2024dcz}. We can also abound in the issue by  looking at the results obtained in Ref.~\cite{Liang:2014kra} and Ref.~\cite{Uchino:2015uha}. In Ref.~\cite{Liang:2014kra} the pseudoscalar-baryon and vector-baryon states in coupled channels were studied using vector exchange in each sector and connecting the two sectors by pion exchange. Tensor exchange is found to be negligible in SU(2) \cite{Dobado:2001rv}, in SU(3) \cite{Ecker:2007us} and also when extended to the charm sector \cite{Abreu:2023wur}. A suitable choice of cutoff parameters to regularize the loops allows one to obtain an acceptable description of the $\Lambda_c(2595)~(1/2^-)$ and $\Lambda_c(2625)~(3/2^-)$ states. Yet, when the same parameters were used to study the hidden charm related states in Ref.~\cite{Uchino:2015uha}, the masses obtained were too low compared with the masses of the $P_c$ states~\cite{LHCb:2019kea}. This gives us an indication that it was the large values chosen for the cutoff parameters what made it posible to get the  $\Lambda_c(2595)$ and  $\Lambda_c(2625)$ states as molecular states.  Conversely, if cutoffs are chosen compatible with the  $P_c$ states, the  $\Lambda_c(2595)$ and  $\Lambda_c(2625)$ states would not be pure molecular.

Other options have been proposed in between which seem more realistic. Indeed, in Ref.~\cite{Luo:2022cun}  it is suggested that the $\Lambda_c(2940)$ could be a $D^* N$ molecular state, something also suggested in Refs.~\cite{He:2006is,Dong:2010xv,He:2010zq,Ortega:2013fta,Wang:2020dhf,Kong:2024scz}. In another recent paper~\cite{Yue:2024paz}  the authors call the attention to the new $\Lambda_c(2910)$, reported recently by the Belle collaboration, and suggest that the $\Lambda_c(2940)$ and $\Lambda_c(2910)$ could correspond to the  $3 / 2^{-},~1/2^{-}$ states of $D^* N$, or viceversa. We shall adopt the same point of view. From the work of Ref.~\cite{Liang:2014kra} for $\Lambda_c^*$ states and the work of Ref.~\cite{Du:2021fmf} (tables 10, 11 with scheme II with explicit pion exchange) for the $P_c(4440)~(3/2^-)$ and $P_c(4457)~(1/2^-)$, we favor the  assignment $\Lambda_c(2940) ~\left(1 / 2^{-}\right), ~\Lambda_c(2910) ~\left(3 / 2^{-}\right)$. The assignment is however different in the pionless theory of Ref.~\cite{Zhang:2023czx}, with $P_c(4440) ~(1 / 2^-), ~P_c(4457) ~(3 / 2^-)$. With a preference for the first assignment, we shall anyway give results for the two scenarios. Then, based
on heavy quark symmetry arguments we must accept a 
$DN$ bound state at a mass around $(2940+2910) / 2$
$-M_{D^*}+M_D \simeq 2783$~MeV. There is actually a
$\Lambda_c^*$ resonance around this energy, the $\Lambda_c(2765)$ with unknown spin and parity,  and we can assume this state
to be the heavy quark spin partner of the $\Lambda_c(2940),~\Lambda_c(2910)$ corresponding to the $D N$ bound state with $1 / 2^{-}$.

\section{Formalism}

\begin{figure}[ht]
    \centering
    \includegraphics[width=0.8\textwidth]{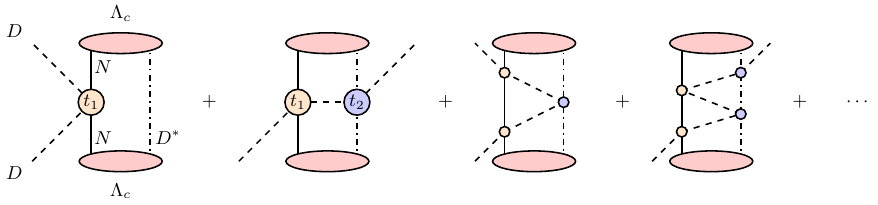}
    \includegraphics[width=0.8\textwidth]{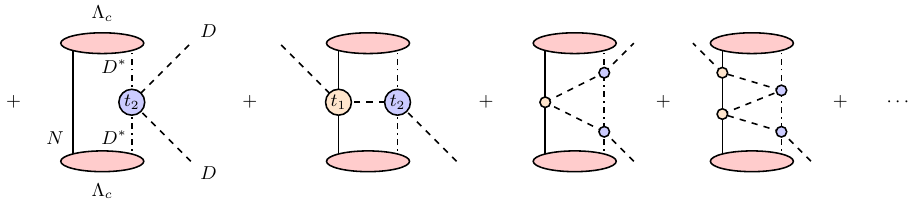}
    \caption{Diagrams corresponding to the  $D$ collision with the $N D^{*}$ cluster. The diagrams starting with collision with $N$ provide the partition function $T_1$. Those starting with $D$ collisions with $D^*$ give rise to $T_2$.}
    \label{Fig1}
\end{figure}

We will assume to have a $D^* N$ cluster either corresponding to  the $\Lambda_c(2940)$ or $\Lambda_c(2910)$, and an external $D$ meson that interacts with this cluster.\footnote{We work in the isospin limit in which the masses of the $D^0$ and $D^+$ mesons, and of the proton and neutron, are the same, and we take for them the
average of their physical masses.} In the FCA approach we have to consider the diagrams shown in Fig.~\ref{Fig1}.
Collecting all diagrams where the external $D$ collides first with $N$ in the partition function $T_1$, and those where the $D$ collides first with $D^*$ into $T_2$, we have formally the  coupled equations
\begin{equation}
\left.
\begin{aligned}
& T_1=t_1+t_1 G_0 T_2 \\
& T_2=t_2+t_2 G_0 T_1 
\end{aligned}
\right\}
\quad  T=T_1+T_2 ,
\end{equation}
where $t_1$ and $t_2$ are the amplitudes for collision of $D$ with $N$ and $D^{*}$ respectively, and $G_0$ corresponds to the $D$ propagator modulated by the wave function of the $D^*N$ system. Yet, we must pay attention to the spin-isospin structure of these amplitudes and to  their normalization, which in the Mandl \& Shaw convention for mesons and baryons that we use are different~\cite{Mandl:1985bg}. We follow closely the steps in Refs.\cite{Roca:2010tf,Xiao:2011rc}. In Ref.~\cite{Xiao:2011rc} the authors studied the $K D N$ system among others.
The normalizations are then the same as here (permuting $1\leftrightarrow 2$) when the $K$ is the third particle. Then we can write immediately
\begin{align}\label{7_1}
  T=\frac{\tilde{t}_1+\tilde{t}_2+2 \tilde{t}_1 \tilde{t}_2 G_0}{1-\tilde{t}_1 \tilde{t}_2 G_0^2} ,
\end{align}
with
\begin{align}
& \tilde{t}_1=t_1(D N),\quad \tilde{t}_2=\frac{1}{2 M_{D^*}} t_2\left(D D^{*}\right),  \\
& G_0=\int \frac{d^3 q}{(2 \pi)^3} F(q) \frac{1}{q^{02}-\vec{q}~^2-M_D^2+i \epsilon}. 
\end{align}
In the previous expressions $q^{0}$ is the energy of  the $D$ in the rest frame of $N D^*$, given by~\cite{Bayar:2022bnc}
\begin{equation}\label{e:D_energy}
    q^0=\frac{s-M_D^2-M_{\Lambda_c^*}^2 }{2 M_{\Lambda_c^*}},
\end{equation}
and $F(q)$ is the form factor of the $N D^*$
state which is given by
\begin{align}\label{8_1}
& F~({ {q }})=\frac{1}{\cal{N}}\int_{|\vec{p}|\leq q_\mathrm{max},~ |\vec{q}-\vec{p}|\leq q_\mathrm{max}}~\frac{d^3 p}{\left(2\pi\right)^3} \frac{1}{M_{\Lambda_c^*}-E_N(\vec{p})-\omega_{D^*}(\vec{p})}~\frac{1}{M_{\Lambda_c^*}-E_N(\vec{p}-\vec{q})-\omega_{D^*}(\vec{p}-\vec{q})},
\end{align}
with $E_N(\vec{p})=\sqrt{M_N^2+\vec{p}~^2}$,~ $\omega_{D^*}(\vec{p})=\sqrt{M_{D^*}^2+\vec{p}~^2}$ and
$\cal{N}$ the normalization factor
\begin{equation}
{\cal{N}} = \int_{|\vec{p}|\leq q_\mathrm{max}}~\frac{d^3 p}{\left(2\pi\right)^3}\left(\frac{1}{M_{\Lambda_c^*}-E_N(\vec{p})-\omega_{D^*}(\vec{p})}\right)^2 . 
\end{equation}
In Eq.~\eqref{8_1} we have introduced a cutoff $q_\mathrm{max}$ which is the regulator used in the loops to get the pole in the $D^*{N}$ interaction. We take $q_\mathrm{max}=600$~MeV suited to get the $P_{cs}$ states in the work of Ref.~\cite{Feijoo:2022rxf}, but we shall see how the results change with $q_\mathrm{max}$.

\subsection{Isospin and spin considerations} 
We will now briefly discuss the isospin structure of the amplitudes. Following Refs.~\cite{Roca:2010tf,Xiao:2011rc,Bayar:2022bnc} and taking into account that $ND^*$ is coupled to  $I=0$, we obtain
\begin{equation}
\begin{aligned}\label{e:t-isospin}
& t_1=\frac{3}{4} t_{DN}^{I=1}+\frac{1}{4} t_{ {DN }}^{I=0}, \\
& t_2=\frac{3}{4} t_{DD^*}^{I=1}+\frac{1}{4} t_{DD^*}^{I=0}.
\end{aligned}
\end{equation}
Since the dominant amplitudes for $DN$ or $DD^*$ are those with $I=0$, we take $ t_1=\frac{1}{4} t_{DN}^{I=0},~ t_2=\frac{1}{4} t_{DD^*}^{I=0}$. We observe that the coefficients $\frac{1}{4}$ weaken the strength of the total $T$ matrix but do not affect much the position of the peak. With respect to the spin structure, since the external $D$ carries none, the total spin in both cases arises from the spin of the cluster, which can be $J= 3 / 2$ or $1 / 2$. Therefore, the spin of the $D N$ and $D D^{*}$ amplitudes is fixed, being $1 / 2$ for the former and $1$ for the latter.
Then, altogether we have
\begin{equation}
\begin{aligned}\label{9_1}
& t_1=\frac{1}{4} t_{ {DN }}^{I=0,~S=1/2}, \\
& t_2=\frac{1}{4} t_{DD^*}^{I=0,~S=1}.
\end{aligned}
\end{equation}

\subsection {\boldmath $D N, ~D D^*$  amplitudes }
%The $ND^*$  interaction is taken into account implicitly, by means of its wave function and, concretely, the form factor of Eq.~(\ref{8_1}). The $DD^*$ amplitude is obtained in Ref.~\cite{}[3] and is given by
%The $ND^*$  interaction is taken into account implicitly, by means of its wave function and, concretely, the form factor of Eq.~(\ref{8_1}). The $DD^*$ amplitude is obtained in Ref.~\cite{Feijoo:2021ppq} and is given by

The $ND^*$  interaction is taken into account implicitly by means of its wave function and, concretely, the form factor of  Eq.~(\ref{8_1}). The $DD^*$ amplitude is obtained in Ref.~\cite{Feijoo:2021ppq} and we approximate it by
\begin{equation}
    t_{DD^*}^{I=0,~S=1}~(s_2) = \frac{(g_{DD^*}^{I=0})^2}{s_2-M_{T_{cc}}^2+i M_{T_{cc}} \Gamma_{T_{cc}}},
\end{equation}
for which we take the empirical values of $M_{T_{cc}}$ and $\Gamma_{T_{cc}}$ of Ref.~\cite{LHCb:2021auc}
\begin{equation}
M_{T_{cc}}=M_{D^0D^{*+}}-360~ \mathrm{keV},\qquad  \Gamma_{T_{cc}}=48~ \mathrm{keV}.
\end{equation}
As to the $D D^*$  coupling, we obtain it from Ref.~\cite{Feijoo:2021ppq}, as follows. Consider the $I=0$ wave function\footnote{Note that this definition is consistent with the $\left( D^{+}, -D^{0} \right)$,  $\left( D^{*+}, -D^{*0} \right)$ isospin doublet definition.}
\be
|DD^{*},~I=0\rangle=-\frac{1}{\sqrt{2}}\left( |D^+D^{*0}\rangle- |D^0D^{*+}\rangle\right).
\ee
From here, we obtain
\be
\langle DD^{*},~I=0~|~t~|~DD^{*},~I=0\rangle=\frac{1}{{2}}\left(t_{11}-t_{12}-t_{21}+t_{22}\right),
\ee
with $1\equiv D^+D^{*0},~ 2\equiv D^0D^{*+}$. By taking the amplitudes as 
$ t_{ij}=g_i g_j /\left(s-M_{T_{cc}}^2+i M_{T_{cc}}\Gamma_{T_{cc}}\right)$
one gets:
\be
\left(g_{DD^*}^{I=0}\right)^2=\frac{1}{2}\left(g_{D^+D^{*0}}^2+g_{D^0D^{*+}}^2-2g_{D^+D^{*0}}g_{D^0D^{*+}}\right).
\ee
Then using the values from Ref.~\cite{Feijoo:2021ppq}
\be
g_{D^+D^{*0}}=-3921~\mathrm{MeV}, \qquad g_{D^0D^{*+}}=3658~\mathrm{MeV},
\ee
we finally obtain
\be
g_{DD^{*}}^{I=0}=5359~\text { MeV }.
\ee
We also need the $DN$ amplitude, which we write as
\begin{equation}\label{10_1}
     t_{DN}^{I=0} = \frac{g_{DN}^2}{\sqrt{s_1}-M_{\Lambda_c({DN})}+i\Gamma_{\Lambda_c({DN})}/2} , 
\end{equation}
with $M_{\Lambda_c({DN})}$ and $\Gamma_{\Lambda_c({DN})} $ corresponding to $\Lambda_c(2765)$,
\be
M_{\Lambda_c({DN})}= 2766.6~ \text { MeV },\qquad \Gamma_{\Lambda_c({DN})}= 50~\text { MeV }.
\ee

We do not make a model for the $\Lambda_c(2765)$ and then we get $g_{DN}$ from $S-$wave Weinberg's compositeness condition \cite{Weinberg:1965zz} using the formula of Ref.~\cite{Gamermann:2009uq}  suited to the normalization of Eq.~(\ref{10_1}) as\footnote{We note that this resonance might have also a molecular $\Sigma_c \pi$ component, as discussed in Ref~\cite{Nieves:2024dcz}. However, for this exploratory study, we neither consider this possibility nor the influence of the quark-model radial $2S$ state.}
\begin{equation}\label{e:coupling}
g_{DN}^2=\frac{M_{\Lambda_c(DN)}}{4 M_N \mu} 16 \pi \gamma, \qquad \gamma=\sqrt{2 \mu B},
\end{equation}
with $\mu$ the reduced mass of $DN$ and $B$ the binding of $\Lambda_c(2765)$ with respect to the $DN$ threshold. We
obtain
\be
g_{DN }=3.70,
\ee
which is  in line with the couplings obtained for the $\Lambda_c^*$  states in Ref.~\cite{Liang:2014kra}.
{It is interesting to note that the scattering length that we get from this model is $a = (1.25-i0.73)$ fm. The real part agrees within errors with the value obtained in Ref.~\cite{Sakai:2020psu} from the analysis of the $pD^0$ mass distribution around threshold in the $\Lambda_b \rightarrow \pi^- pD^0$ decay, after changing the sign due to a different definition of $a$ (the imaginary part is omitted in Ref.~\cite{Sakai:2020psu}).}

The last point we need to complete the formalism is the value of the argument of the $DN$ and $DD^*$ amplitudes $s_1,~ s_2$, which are given by~
\cite{Roca:2010tf}
%$$
%\begin{aligned}
%    s_1&= M_D^2+ \left(M_N \frac{M_{\Lambda_i}}{M_N+M_{D^\ast}}\right)^2+ 2 \left(M_N \frac{M_{\Lambda_i}}{M_N+M_{D^\ast}}\right) \frac{s-M_D^2-M_{\Lambda_i}^2}{2M_{\Lambda_i}}\\
%    s_2&= M_D^2+ \left(M_{D^\ast} \frac{M_{\Lambda_i}}{M_N+M_{D^\ast}}\right)^2+ 2 \left(M_{D^\ast} \frac{M_{\Lambda_i}}{M_N+M_{D^\ast}}\right) \frac{s-M_D^2-M_{\Lambda_i}^2}{2M_{\Lambda_i}}
%\end{aligned}
%$$
\be
\begin{aligned}
    s_1&= M_D^2+ \left(\xi M_N\right)^2+ 2 \xi M_N q^0,\\
    s_2&= M_D^2+ \left(\xi M_{D^\ast}\right)^2+ 2 \xi M_{D^\ast} q^0,
\end{aligned}
\ee
where $\xi = M_{\Lambda_c^*}/(M_N+M_{D^\ast})$ is a correction factor that takes into account the ``off-shellness'' of the $N$ and $D^*$ inside of the corresponding $\Lambda_c^*$ resonance, and $q^0$ is the energy of the $D$ in the rest frame of the $ND^*$ pair, as given in Eq.~\eqref{e:D_energy}.\footnote{These formulas for $s_1$ and $s_2$ differ slightly when compared to the ones used in Refs.~\cite{Xiao:2011rc,Yamagata-Sekihara:2010muv}. However, they are shown to be exactly equivalent in the limit in which $M_{\Lambda_c^*}=M_N+M_{D^\ast}$.}

%\sj{Can you see what comes out using  Eq. (20) of Bryar PRD $84,034037 $ ?} \Vic{Different formula! Only equivalent in the limit in which $M_{\Lambda_i}=M_N+M_{D^*}$}

\subsection {\boldmath The $D^*(ND)$ system}\label{sec:NDcluster}

Alternatively we can also assume the $ND$ system, which is relatively well bound, as a cluster and the $D^*$ as the external particle. The formalism is identical to the former one, but we only have $J^P=1 / 2^{-}$ for the $ND$ cluster. Conversely, we still have $1 / 2^{+},~ 3 / 2^{+}$ states from the $D^{*} N$ interaction in $J^{P}=1 / 2^{-},~3 / 2^{-}$. Thus we will have now
\be
t_1=\frac{1}{4} t_{D^* N}^{I=0, S=1 / 2,~3 / 2},\quad \text { for } 1 / 2^{+},\,  3 / 2^{+} \text{ final three body states,}
\ee
where
\begin{align}
    t_{D^* N}^{I=0, S=1 / 2} = \frac{(g_{D^*N}^{S=1/2})^2}{\sqrt{s_1}-M_{\Lambda_c(2940)}+i\Gamma_{\Lambda_c(2940)}/2} ,\\
    t_{D^* N}^{I=0, S=3 / 2} = \frac{(g_{D^*N}^{S=3/2})^2}{\sqrt{s_1}-M_{\Lambda_c(2910)}+i\Gamma_{\Lambda_c(2910)}/2} ,
\end{align}
and $t_2$ the same as before in Eq~\eqref{9_1}. In order to compute the couplings of the $D^* N$ pair to the corresponding $\Lambda_c^*$ states, we make use again of Eq.~\eqref{e:coupling} and obtain
\be
g_{D^*N}^{S=1/2} = 2.63 ~ , \quad g_{D^*N}^{S=3/2} = 3.71 .
\ee
These values are also in line with the results in Ref.~\cite{Liang:2014kra}. We also have to replace $M_{\Lambda_c^*}$ in Eq~\eqref{8_1} by the mass of the $\Lambda_c(2765)$ and $\omega_{D^*}(p)$ by $\omega_D(p)$. The formulas for $q^{0},~ s_1,~ s_2$ are now finally modified by permuting $D \leftrightarrow D^*$ and using the mass of the $\Lambda_c(2765)$ for the cluster.

\section{Results}
\begin{figure}[H]
    \centering
    \includegraphics[width=0.55\textwidth]{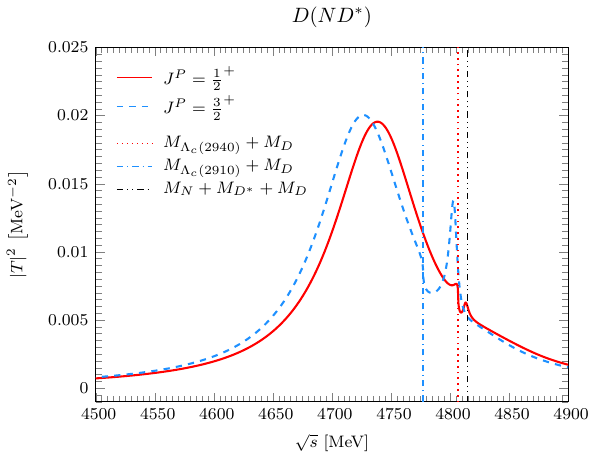}
    \caption{$|T|^2$ as a function of the total energy of the three-body system, $\sqrt{s}$, for $q_\mathrm{max}=600$ MeV, assuming the $ND^*$ cluster pair to be bound into the $\Lambda_c(2940)$ and $\Lambda_c(2910)$ for spin $1/2$ and $3/2$, respectively.}
    \label{fig:AmplitudeD*Ncluster}
\end{figure}

In Fig.~\ref{fig:AmplitudeD*Ncluster} we show the results for $|T|^2$ assuming the $\Lambda_c(2940)$ to be a $J^P=1/2^-$ state and the $\Lambda_c(2910)$ to be $3/2^-$. We obtain two states, one with total $J^P=1/2^+$ and another state with $3/2^+$. With respect to the thresholds of $2940 \ \mathrm{MeV}+M_D$ and $2910 \ \mathrm{MeV}+M_D$, the states are bound by about $70$ and $50$ MeV, respectively. We can see that we get peaks for $|T|^2$ with widths that correspond to the width of the three-body system. The widths obtained are of the order of $90$ MeV. If we make the opposite spin assumption for the $\Lambda_c(2940)$ and $(2910)$ states, we obtain the same results but with the total spins exchanged  (red, solid line $\leftrightarrow$ blue, dashed line in Fig.~\ref{fig:AmplitudeD*Ncluster}).

It is interesting to make an observation with respect to the results obtained. The peaks obtained correspond basically to the structure of $\tilde t_1$ ($DN$ amplitude) in Eq.~\eqref{7_1}, and the terms with $\tilde t_2$ in the numerator and denominator of Eq.~\eqref{7_1} do not play much of a role. This is understandable since the strength of $\tilde t_2$ is very small compared to $\tilde t_1$. Indeed, the $\Lambda_c(2765)$ is bound by about $40$ MeV while the $T_{cc}$ is only bound by $360$ keV, indicating that the $DN$ interaction is much stronger than the $DD^*$
one. The picture is that the $N$ is acting as a glue to which both the $D$ and $D^*$ stick, but with the $DD^*$ interaction being so weak, this is the realistic picture emerging for this three-body system.

The narrow peak seen in the $J^P=3/2^+$ case between thresholds comes from the $T_{cc}$ amplitude. There are other structures close to the different thresholds which can be identified as threshold effects. They could also correspond to less bound
states of the system. Such structures also show up in the $D^*D^*D^*$ bound state~\cite{Bayar:2022bnc} and they appear as excited states over the fundamental three body state when using the ladder amplitude formalism \cite{Hansen:2015zga,Jackura:2020bsk,Dawid:2023jrj} in relationship to the Efimov effect in Ref.~\cite{Ortega:2024ecy}.

\begin{figure}[ht]
    \centering
    \includegraphics[width=0.55\textwidth]{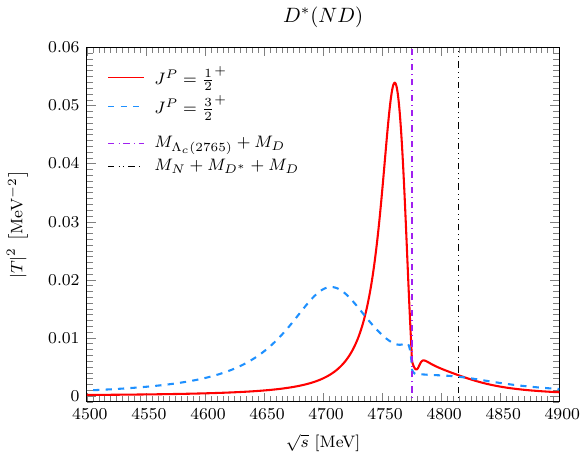}
    \caption{{$|T|^2$ as a function of the total energy of the three-body system, $\sqrt{s}$, for $q_\mathrm{max}=600$ MeV, assuming the $ND$ cluster pair to be bound into the $\Lambda_c(2765)$.}}
    \label{fig:AmplitudeDNcluster}
\end{figure}
As an alternative picture, we have also considered the  scenario in which the $ND$ system is  the cluster
and is bound into the $\Lambda_c(2765)$ (assuming it has $J^P=1/2^-$), and there is an external $D^*$ interacting with it, as explained in Sec.~\ref{sec:NDcluster}. The results in this picture are shown in Fig.~\ref{fig:AmplitudeDNcluster}. In this case, we observe a binding energy with respect to the $2765 \ \mathrm{MeV}+M_{D^*}$ threshold of $15$ MeV for the $1/2^+$ state, and of $70$ MeV for the $3/2^+$ state. When comparing it to the plot of Fig.~\ref{fig:AmplitudeD*Ncluster}, the most notable change is for the $J^P=1/2^+$ amplitude, which is now considerably narrower, with a width of about $30$ MeV. This can be understood because of the smaller binding energy that the $1/2^+$ state   has now with respect to the $\Lambda_c(2765) D^*$ threshold. 
For completeness, the absolute masses obtained from the peak positions in both pictures are shown in table~\ref{table1}.
\begin{table}[H]
\centering
 \caption{Masses of the bound states $DND^*$ in $\mathrm{MeV}$. On the tabular on the left-hand side we show the results obtained when assuming the $ND^*$ pair to be bound, while the tabular on the right contains the corresponding results when considering the $ND$ pair to be bound.}\label{table1}
\setlength{\tabcolsep}{6pt}
\begin{tabular}{ccc}
\multicolumn{3}{c}{$D(ND^*)$}\\ \hline
$q_\mathrm{max}$ [MeV] & $J^P$ & mass [MeV] \\ \hline \hline
\multirow{2}{*}{600} & $1/2^+$ & $4738.6$ \\
& $3/2^-$ & $4726.5$ \\ \hline
\multirow{2}{*}{650} & $1/2^+$ & $4738.6$ \\
& $3/2^-$ & $4726.6$ \\
\end{tabular}
\hspace{1cm}
\begin{tabular}{ccc}
\multicolumn{3}{c}{$D^*(ND)$}\\ \hline
$q_\mathrm{max}$ [MeV] & $J^P$ & mass [MeV] \\ \hline \hline
\multirow{2}{*}{600} & $1/2^+$ & $4760.6$ \\
& $3/2^-$ & $4706.8$ \\ \hline
\multirow{2}{*}{650} & $1/2^+$ & $4759.6$ \\
& $3/2^-$ & $4706.9$ \\
\end{tabular}
\end{table}
As we can see, the values of the masses  are very similar in both scenarios, with differences of around $20$ MeV. This change on their values can be taken as an estimate of the uncertainties associated with the FCA formalism. We also show the variation of these masses with a change on the value of $q_\mathrm{max}$, which we find negligible. Yet, it
is interesting to note that some of the small structures close to threshold disappear
with the use of a larger $q_\mathrm{max}$. {In particular, the narrow peak seen in Fig.~\ref{fig:AmplitudeD*Ncluster} in the $3/2^+$ amplitude between thresholds is notably affected and for $q_\mathrm{max}=700$ MeV this structure has almost dissapeared.} This feature reminds one of the findings of Ref.~\cite{Ortega:2024ecy}
where the excited states obtained for the $D^*D^*D^*$ system also disappear with larger
values of the cutoff.

\subsection{\boldmath Consideration of $DN$ isospin $I=1$}
We come back to our favored configuration of $D$ scattering with the $N D^*$ cluster. In Eq.~\eqref{e:t-isospin}, $t_1$ had a combination of $I=0$ and $I=1$ contributions and we took $I=0$ to be the dominant one. The dominance of the $I=0$ interaction is rather common, and in all molecular pictures discussed above the $DN$ interaction for $I=1$ is much weaker than for $I=0$. Yet, in most of these pictures an $I=1$ state appears, much less bound than the $I=0$ one (see for instance Refs.~\cite{Mizutani:2006vq,Liang:2014kra}). This situation has led the authors of Refs.~\cite{Luo:2022cun,Sakai:2020psu} to suggest that the $DN$ interaction in $I=1$ is responsible for the state $\Sigma_c(2800)$, close to the $DN$ threshold. We shall adopt the same view and then the $DN$ $I=1$ amplitude is given by
\be
t^{I=1}_{DN} = \frac{(g^{I=1}_{DN})^2}{\sqrt{s_1}- M_{\Sigma_c}+ i \Gamma_{\Sigma_c}/2},
\ee
with $g^{I=1}_{DN}$ calculated via Eq.~(\ref{e:coupling}), for which we get $g^{I=1}_{DN}=2.51$. Once again, we can obtain the $DN$ scattering length and we get $a= (0.20 -i 0.84)$ fm. The imaginary part is compatible with the results of Ref.~\cite{Sakai:2020psu}, and the real part has the same sign but is smaller in size. However it is more similar to results obtained in Refs.~\cite{Hofmann:2005sw,Garcia-Recio:2008rjt}. 

\begin{figure}[ht]
    \centering
    \includegraphics[width=0.55\textwidth]{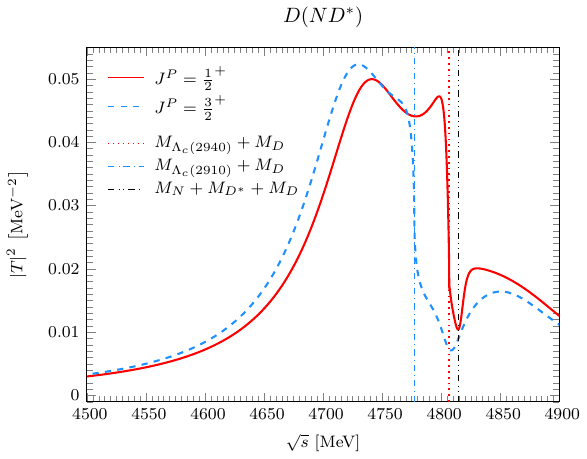}
    \caption{Same as Fig.~\ref{fig:AmplitudeD*Ncluster}, considering the $DN$ $I=1$ amplitude in Eq.~\eqref{e:t-isospin}.}
    \label{fig:AmplitudeD*Ncluster-DNIsovectorAmplitude}
\end{figure}

We then use $t_1$ of Eq.~\eqref{e:t-isospin} rather than the $I=0$ truncated amplitude of Eq.~\eqref{9_1}. We show the new results in Fig.~\ref{fig:AmplitudeD*Ncluster-DNIsovectorAmplitude}. As we can see, the structure for $|T|^2$ concerning the peaks of Fig.~\ref{fig:AmplitudeD*Ncluster} has not changed. Yet, a new structure has appeared in between the two $\Lambda^* D$ thresholds, more prominent for $J^P=1/2^+$. We also note that the earlier small structures around thresholds in Fig.~\ref{fig:AmplitudeD*Ncluster} have disappeared in favor of the new structure produced by the $I=1$ $DN$ amplitude. 

\section{Conclusions} 

%We engage in a discussion around the $\Lambda_c(2595)$ and $\Lambda_c(2625)$ and its possible relation to the $DN$ and $D^*N$ bound states. While this option had some support in the past, the new information on the $P_c, ~P_{cs}, ~T_{cc}$ states has shown contradictions in this picture and stressed the relevance of conventional three quark state contributions to their structure. Parallelly, 

We have carried out a study of the possible bound states of the $DND^*$ system,
for which we use the FCA to make the evaluation of the three body scattering matrix. This requires the choice of a bound two body cluster, for which we take the $ND^*$ system, and a third particle, a $D$ meson, that scatters with the components of the cluster.
The finding of new $\Lambda_c^*$ states, in particular the $\Lambda_c(2940)$ and $\Lambda_c(2910)$, and their proximity to the $D^*N$ threshold, has led many authors to propose that these are the states that should be associated to possible $D^*N$ bound states, with $J^P=1/2^-, 3/2^-$. Certainly the proximity of a state to a meson-baryon threshold favors its association to a bound state of this channel~\cite{Dong:2021juy,Guo:2017jvc}, although it is not necessarily the case as shown in Refs.~\cite{Dai:2023kwv,Song:2023pdq,Li:2023pjx}, at the price of having abnormally small scattering lengths and large effective ranges for the channels involved. Then, the use of heavy quark symmetry leads us to suggest that the $\Lambda_c(2765)$ corresponds to the related bound $DN$ state. 
%From this new perspective we undertake the task of looking for possible bound states of the $D D^* N$ system for which we use the FCA to make the evaluation of the three body scattering matrix. We start from a cluster of $N D^*$, which is bound in this picture, and allow an external $D$ to collide with this cluster. 

Alternatively, we also assume the $N D$ system to be the cluster and allow the $D^*$ to collide with the cluster. We find bound states in $J^P=1/2^+,3/2^+$ with respect to the $\Lambda_c(2765) D$ or $\Lambda_c(2765) D^*$ thresholds, qualitatively similar in both cases.  We also show that the consideration of the $I=1$ $DN$ amplitude, in addition to the $I=0$ one, adds a new structure between the two $\Lambda^* D$ thresholds, leaving the $I=0$ peaks unchanged. Although we favor the $D (ND^*)$ picture, the different masses obtained from the two pictures discussed above can be thought of as a measure of the uncertainties in our approach, from where we conclude the unavoidable existence of these bound states, under the reasonable assumptions made for the $D^* N$  and $D N$ bound states. The observation of such states would shed light on the assumptions made concerning the $D^* N$ and $D N$ bound states, helping clarify the issue of the $DN$ and $D^*N$ interaction.  

\acknowledgments

We would like to thank Pablo Ortega for useful discussion. This work is partly supported by the Spanish Ministerio de Economia y Competitividad (MINECO) and European FEDER funds under contract No. PID2020-112777GB-I00, and by Generalitat Valenciana under contract PROMETEO/2020/023. This project has received funding from the European Union Horizon 2020 research and innovation programme under the program H2020-INFRAIA-2018-1, grant agreement No. 824093 of the STRONG-2020 project. This work of J. S. is partly supported by the National Natural Science Foundation of China under Grants No. 12247108 and the China Postdoctoral Science Foundation under Grant No. 2022M720360 and No. 2022M720359. M. A. and V. M. are supported through Generalitat Valenciana (GVA) Grants No. CIDEGENT/2020/002 and ACIF/2021/290, respectively.
This work is partly supported
by the National Natural Science Foundation of China
(NSFC) under Grants No. 12365019 and No. 11975083,
and by the Central Government Guidance Funds for
Local Scientific and Technological Development, China
(No. Guike ZY22096024).

\bibliography{refs.bib}
\end{document}